\documentclass[pra,aps,showpacs,twocolumn,floatfix]{revtex4}
\usepackage{graphicx}
\usepackage[ansinew]{inputenc}
\usepackage{array}
\usepackage{color}
\usepackage{amsmath}
\usepackage{amsxtra}
\usepackage{amstext}
\usepackage{amssymb}
\usepackage{latexsym}
\usepackage{dsfont}
\usepackage{verbatim}
\usepackage{comment}

\renewcommand{\vec}[1]{{\bf{#1}}}

\begin{document}

\title{Slow-light probe of Fermi pairing through an atom-molecule dark state}
\author{H. Jing$^{1,2}$, Y. Deng$^1$, and P. Meystre$^2$}
\affiliation{$^1$Department of Physics, Henan Normal University,
Xinxiang 453007, China\\
$^2$B2 Institute, Department of Physics and College of Optical
Sciences, The University of Arizona, Tucson, Arizona 85721}
\date{\today}

\begin{abstract}
We consider the two-color photooassociation of a quantum degenerate atomic gas into ground-state diatomic molecules via a molecular dark state. This process can be described in terms of a lambda level scheme that is formally analogous to the situation in electromagnetically-induced transparency (EIT) in atomic systems, and therefore can result in slow light propagation. We show that the group velocity of the light field depends explicitly on whether the atoms are bosons or fermions, as well as on the existence or absence of a pairing gap in the case of fermions, so that the measurement of the group velocity realizes a non-destructive diagnosis of the atomic state and the pairing gap.

\end{abstract}

\maketitle

\section{INTRODUCTION}

Degenerate atomic Fermi gases have attracted much interest in recent years, well past the confines of traditional atomic, molecular and optical (AMO) physics~\cite{BCS}. The existence of correlated Fermi pairs results in a number of effects that can be explored particularly well in these systems, due in particular to the control of two-body interactions provided by Feshbach resonances. These include detailed studies of the crossover from Bardeen-Cooper-Schrieffer (BCS) superfluidity to Bose-Einstein condensation (BCS)~\cite{BCS}, of crystalline and supersolid phases~\cite{solid}, as well as spin-charge separation or spin drag~\cite{drag}, to mention by a few examples. However, in absence of any obvious change of density profile, the detection of Fermi pairing is challenging, in sharp contrast to the familiar BEC transition of bosons. A long-standing goal remains therefore to develop methods to efficiently detect the pairing signature of fermionic systems and other related exotic phases. Approaches toward this goal have focused on the measurement of atomic density-density correlations via the resonant or non-resonant optical response of the fermionic atoms~\cite{laser}, including methods of radio-frequency spectroscopy~\cite{Chin}, photoemission spectroscopy~\cite{emi}, and Raman spectroscopy~\cite{cote}. Alternative methods, like scanning tunneling microscopy~\cite{inf} or acoustic attenuation~\cite{molprob}, are also actively pursued.

In parallel to these developments, rapid experimental advances have resulted in the coherent formation of ultracold molecules from Bose or Fermi atoms~\cite{mol}. The stable formation of diatomic molecules from laser-cooled alkali atoms has been achieved by using magnetic Feshbach resonances and optical photoassociation (PA) techniques. By applying an all-optical PA method, molecules associated from ultracold atoms can be successfully transferred into their rovibrational ground state~\cite{mol2}.

A key component of the two-color PA method is the existence of an atom-molecule dark state, as first demonstrated by Winkler~$\it{et~al.}$~\cite{AMDS}. The underlying quantum interference and slow light propagation were also observed for ultracold sodium atoms by Turner~$\it{et~al.}$~\cite{Tur}, hinting at the possibility to study the quantum control of light through cold reactions~\cite{mol,mol2,AMDS,Tur,HJ}, quantum state transfer from light to molecules \cite{HJ,Letok}, as well as high-precision diagnostics of Fermi gases via PA spectroscopy~\cite{cote}.

In this paper we show that the slow light propagation associated with the existence of that dark state provides a relatively simple nondestructive probe of Fermi pairing, without the need for additional excitations (atom-to-atom, atom-ion-to-molecule, or molecule-to-molecule) or for laser imaging of the populations of transferred particles. This proposed method finds its motivation in a previous work ~\cite{Meiser} which showed that the statistical properties of the molecular field formed from ultracold atoms depends strongly on the statistical properties of these atoms. In particular, it was found that for short times, the number of molecules created scales as the square $N^2$ of the number of atoms in case of an atomic Bose-Einstein condensate, but as $N$ for a normal Fermi gas at zero temperature, a manifestation of the independence of all atomic pairs in that case. For a paired Fermi gas, the situation is intermediate between these two extremes: the molecules are formed at a higher rate than for a normal Fermi gas, and the maximum number of molecules is larger, approaching the BEC situation for strong pairing.

The main result of the present analysis is that a related situation occurs when considering the dark-state propagation of a photoassociating light field: in contrast to the case where photoassociation originates from a condensate of bosonic atoms, and where the inverse group velocity $v_g^{-1}$ of the light field is known to scale as $N^2$, we find that for a normal Fermi gas at $T=0$ it scales as $N$. A paired Fermi system represents an intermediate situation, as was the case in Ref.~\cite{Meiser}. It follows that the group velocity is a direct measure of the pairing gap $\Delta.$ This simple all-optical method is also expected to prove useful in probing e.g. polaron-to-molecule transitions and atom-molecule vortex states~\cite{polaron} by photoassociating a spin-imbalanced or a rotating Fermi gas. We remark that this proposal involves the use of tunable atom-molecule interactions and as such is fundamentally different from approaches based on single-atom excitations~\cite{laser,f-eit}.

The paper is organized as follows. Section II describes our model and calculates the slow light group velocity of a quantized optical field that propagates in a normal Fermi gas and helps photoassociating atoms into molecules via a dark state intermediate level. Section III evaluates the effect of a Fermi pairing gap on that velocity and shows that it depends strongly on the magnitude of the gap. Finally Section IV is a
conclusion and outlook.

\section{Normal Fermi Gas}

We first consider the two-color photoassociation of a homogeneous, normal degenerate Fermi gas with no pairing. The entrance channel atoms, the intermediate state $|m\rangle$ and the closed channel bosonic molecules are characterized by the annihilation operators $\hat{c}_{\vec k\sigma}$, $\hat{m}_{\vec{k+k'}}$ and $\hat{a}$, respectively, where $\vec k$ and $\vec k'$ are wave numbers and $\sigma$ labels the fermionic spin. We assume that the PA between atomic pairs and excited molecules in state $|m\rangle$ is driven by an optical field that is treated quantum mechanically at that point, and the field that drives the molecules to their ground state $|g\rangle$ is classical, with Rabi frequency $\Omega(t)$ (see Fig.~1).

\begin{figure}[ht]
\includegraphics[width=0.85\columnwidth]{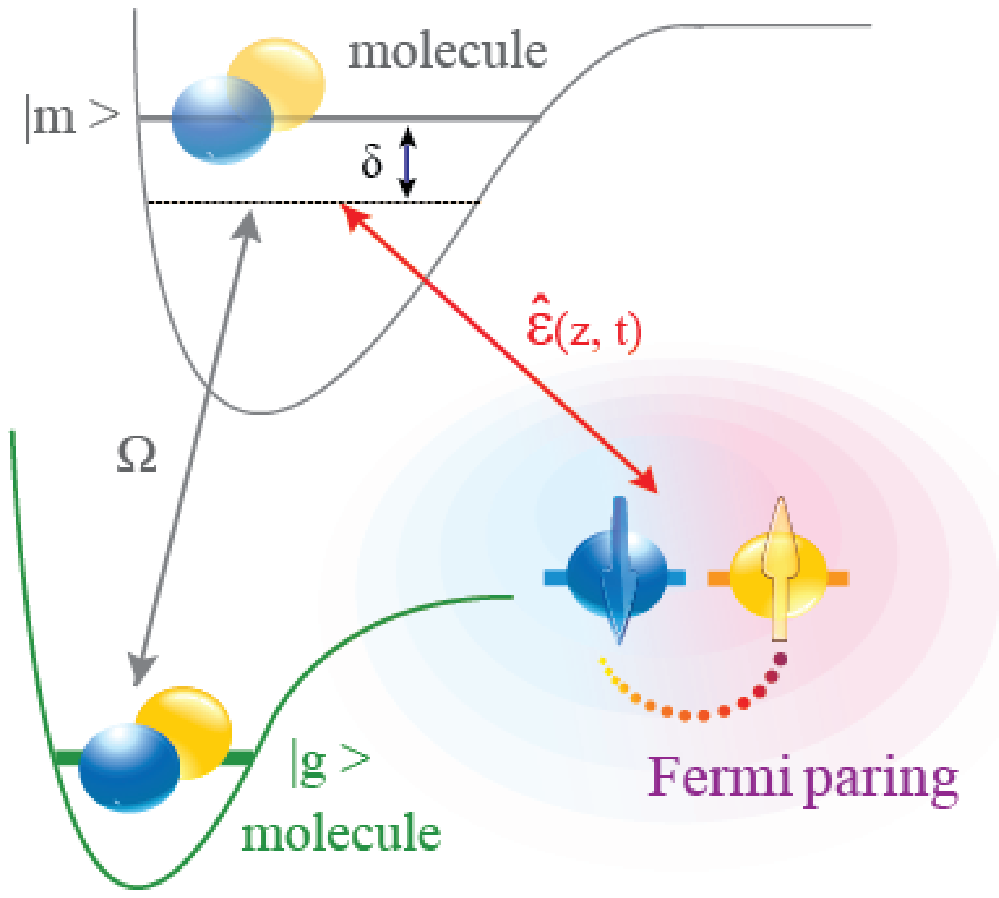}
\caption{Schematic of two-color PA in an ultracold
degenerate Fermi gas with or without Cooper pairing.}
\label{fig2}\end{figure}

At the simplest level the Hamiltonian of this system can be expressed as
($\hbar =1$)
\begin{eqnarray}
\hat{H}&=&\sum_{\vec{k},\sigma} \frac{\epsilon_\vec{k}}{2}\hat{c}_{\vec{k}\sigma}^\dagger \hat{c}_{\vec{k}\sigma}+ g\sum_{\vec{k},\vec{k'}}\left(\hat{\cal E} \hat{m}_{\vec{k+k'}}^\dagger \hat{c}_{\vec{k}\uparrow}\hat{c}_{\vec{k'}\downarrow} + \text{h.c.}\right)
\nonumber \\
&+&\sum_\vec{k, k'}\left [ \delta\hat{m}_{\vec{k+k'}}^\dagger \hat{m}_{\vec{k+k'}}+ \Omega( \hat{a} \hat{m}_{\vec{k+k'}}^\dagger+ \text{h.c.})\right ] , \label{full_hamiltonian}
\end{eqnarray}
where $g$ is the atom-molecule coupling constant, $\delta$ is the detuning between the frequency of the quantized photoassociation field and the frequency difference between the atomic fermions and the molecular state $|m\rangle$ -- we neglect the dispersion in fermionic energies $\epsilon_{\vec{k}}$ for simplicity --  and $\Omega(t)$ is the Rabi frequency of the classical field, taken to be real without lack of generality. The $s$-wave collisions between fermionic atoms, between molecules, and between atoms and molecules are ignored for a dilute gas.

For simplicity, we restrict ourselves to the association of atom pairs with opposite momenta ($\vec {k=-k'}$) and opposite spin, in which case the intermediate molecules can be also described in terms
a single-mode bosonic field when concentrating on short-time dynamics, see e.g. Refs.~\cite{Meiser,Holland}. With these simplifying assumption this system is formally analogous to the situation of EIT in atomic lambda systems, and as such can result in slow light propagation.

The quantized optical field $\hat{E}(z, t)$, of carried frequency $\nu$,  is
given by
$$
\hat{E}(z, t)=\sqrt{\frac{\hbar\nu}{2\epsilon_0 L}}\hat{\cal E}(z, t)\exp\left [i\frac{\nu}{c}(z-ct) \right ],
$$
where $L$ is the quantization length. It satisfies the commutation relation
$$
[\hat{E} (z, t),\hat{E} ^\dagger (z',t)] =\frac{\nu}{\epsilon_0}\delta (z-z').
$$
Within the slowly-varying-amplitude approximation, the propagation equation of the field envelope $\hat{\cal E}(z, t)$ is given by
\begin{equation}
\left (\frac{\partial}{\partial t}\!+\!c\frac{\partial}{\partial z} \right )\hat{\cal E}(z, t) = i gL\sum_{\vec{k}}\hat{c}_{\vec{-k}\downarrow}^\dagger(z, t)\hat{c}_{\vec{k}{\uparrow}}^\dagger(z, t)\hat{m}(z, t).
\end{equation}

In the following we consider the regime of  weak excitations, where the atomic population remains essentially undepleted. The initial state of the atom-molecule system is taken as
$$
|\psi(0)\rangle=|F\rangle\otimes |0\rangle_{m}\otimes |0\rangle_a,
$$
where $|0\rangle_m$, and  $|0\rangle_g$ denote the vacuum state for the molecules and
$$
|F \rangle
=\prod_k\hat{c}_{-\vec{k}\downarrow}^\dag\hat{c}_{\vec{k}\uparrow}^\dag|0\rangle,
$$
and the product is taken up to the Fermi surface, a step appropriate for temperatures much below the Fermi temperature~ \cite{Meiser}. Introducing the pseudo-spin operators
\begin{eqnarray}
\hat{s}_{\vec k}^+&=&(\hat{s}_{\vec k}^-)^\dagger=\hat{c}_{\vec{-k}\downarrow}^\dagger\hat{c}_{\vec{k}{\uparrow}}^\dagger, \nonumber \\
\hat{s}_{\vec k}^z&=&\frac{1}{2}\left
(\hat{c}_{\vec{k}\uparrow}^\dagger\hat{c}_{\vec{k}\uparrow}+\hat{c}_{\vec{-k}\downarrow}^\dagger\hat{c}_{\vec{-k}\downarrow}-1
\right ),
\end{eqnarray}
which satisfy the commutation relations
\begin{equation}
[\hat{s}_{\vec k}^+, \hat{s}_{\vec k'}^-] = 2\delta_{\vec {k k'} }\hat{s}_{\vec k}^z, \,\,\,\,\, [\hat{s}_{\vec k}^z, \hat{s}_{\vec k'}^\pm] = \pm \delta_{\vec {k k'}} \hat{s}_{\vec k}^\pm,
\end{equation}
and the collective operators
\begin{eqnarray}
\hat{S}_\pm&=&\sum_{\vec{k}}\hat{s}_{\vec k}^\pm, \nonumber \\
\hat{S}_z&=&\sum_{\vec{k}}\hat{s}_{\vec k}^z = \frac{N}{2}-\hat{a}^\dagger\hat{a}-\hat{m}^\dagger\hat{m}, \nonumber \\
\hat{\vec{S}}^2&=&\hat{S}_+\hat{S}_-+\hat{S}_z(\hat{S}_z-1),
\end{eqnarray}
with the conserved total number of atomic pairs and molecules
\begin{eqnarray}
N&=&\sum_k \left (\hat{c}_{\vec{k}\uparrow}^\dag\hat{c}_{\vec{k}\uparrow}+\hat{c}_{-\vec{k}\downarrow}^\dag\hat{c}_{-\vec{k}\downarrow} \right )/2+(\hat{a}^\dagger\hat{a}+\hat{m}^\dagger\hat{m}) \nonumber \\
&=&(\hat{S}_z+N/2)+(\hat{a}^\dagger\hat{a}+\hat{m}^\dagger\hat{m}),
\end{eqnarray}
yields for the Hamiltonian $\hat{H}_\mathcal{N}$ the simplified form
\begin{equation}
\hat{H}=\sum_{\vec k} \epsilon_\vec{k}\hat{s}^z_{\vec k}+\delta\hat{m}^\dagger\hat{m}+ \left(g\hat{\cal E}\hat{m}^\dagger\hat{S}_{-}+\Omega\hat{m}^\dagger \hat{a}+{\rm h.c.}\right).
\end{equation}
The resulting Heisenberg equations of motion are, by approximating
all $\epsilon_{\vec k}$'s as the Fermi energy $\epsilon_F$,
\begin{eqnarray}
i\frac{d{\hat{S}}_z}{dt} &=& g\hat{\cal E}^\dagger\hat{m}\hat{S}_+-g\hat{\cal E}\hat{m}^\dagger\hat{S}_-,\nonumber \\
i\frac{d{\hat{S}}_-}{dt} &=&\epsilon_F\hat{S}_-- 2g\hat{\cal E}^\dagger\hat{m}\hat{S}_z,\nonumber \\
i\frac{d{\hat{S}}_+}{dt} &=&-\epsilon_F\hat{S}_-+2g\hat{\cal E}\hat{m}^\dagger\hat{S}_z,\nonumber \\
i\frac{d\hat{m}}{dt} &=&g\hat{\cal E}\hat{S}_-+\delta\hat{m}+\Omega\hat{a},\nonumber \\
i\frac{d\hat{a}}{dt} &=&\Omega\hat{m},\nonumber \\
i\frac{d\hat{\cal E}}{dt}&=&g\hat{m}\hat{S}_-.
\end{eqnarray}
In the following we consider the resonant situation $\delta = 0$ and the limit of weak excitations. By setting $d\hat{m}/dt\rightarrow 0$, we have then in the lowest nonvanishing order of the excited molecular state~\cite{Lukin, HJ},
\begin{eqnarray}
\hat{a}&=&-{(g/{\Omega})\hat{\cal E}}\hat{S}_-, \nonumber \\
\hat{m}&=&-i(g/{\Omega}){\hat{S}_-}\frac{\partial}{\partial t}(\frac{\hat{\cal E}}{\Omega}).
\label{equation}
\end{eqnarray}

The propagation of the field $\hat {\cal E}(z,t)$  is then governed by the equation
\begin{equation}
\left (\frac{\partial}{\partial t}+c\frac{\partial}{\partial z}\right )\hat{\cal E}(z, t)=- \frac{g^2 L N}{\Omega}\frac{\partial}{\partial t}\left (\frac{\hat{\cal E}}{\Omega}\right ),
\label{propagation 1}
\end{equation}
where we have used
\begin{equation}
\hat{\vec{S}}^2|F\rangle=S(S+1)|F\rangle=\frac{N}{2}\left(\frac{N}{2}+1\right)|F\rangle,
\end{equation}
and the weak excitation approximation
\begin{equation}
\langle \hat{S}_+\hat{S}_- \rangle = (-n_a^2+n_aN-n_a)+N \sim N.
\end{equation}
Equation~(\ref{propagation 1}) can be recast as
\begin{eqnarray}
\left (\frac{\partial}{\partial
t}+\frac{c}{1+\beta_f}\frac{\partial}{\partial z} \right )\hat{\cal
E}(z, t)=\frac{\beta_f}{1+\beta_f}\left (
\frac{1}{\Omega}\frac{\partial \Omega}{\partial t} \right )
\hat{\cal E}. \label{propagation 2}
\end{eqnarray}
where
\begin{equation}
\beta_f \equiv \frac{g^2 L N}{\Omega^2}.
\label{betaf}
\end{equation}
That is, the group velocity of the field $\hat {\cal E}(z,t)$ is
\begin{equation}
v_{g}=\frac{c}{1+\beta_f} = c \cos^2\theta,
\end{equation}
with
\begin{equation}
\theta=\tan^{-1}(g\sqrt{LN}/\Omega).
\end{equation}
As mentioned in the introduction, the scaling of $\beta_f$ with $N$ should be contrasted with the situation for a pure condensate of bosonic atoms, in which case~\cite{HJ}
\begin{equation}
\beta_f \rightarrow \beta_b=\frac{g^2 L N^2}{\Omega^2} = N \beta_f.
\end{equation}
As was the case in the analysis of molecule formation of Ref.~\cite{Meiser}, this difference is due to the fact that for a Bose-Einstein condensate the photoassociation is a collective atomic effect, while in a normal Fermi gas the atom pairs act independently from each other.

We remark that the form of $v_g$ is independent of whether the field ${\hat {\cal E}}(z,t)$ is treated classically or quantum mechanically (see the related experiment of Ref.~\cite{Tur}). The quantized description used here is primarily to facilitate a direct comparison with the bosonic atom-molecule system of Ref.~\cite{HJ}. Note however that Eqs.~(\ref{equation}) shows that the statistical properties of the closed-channel molecules are determined by the states of both the  optical field and the Fermi atoms, hinting at the possibility  of quantum control of the closed-channel molecules, e.g. by applying a squeezed PA field~\cite{HJ}.

The next section expands these considerations to the case of a paired Fermi gas, which is then expected to represent an intermediate situation between these two extremes. We show that this is indeed the case, and as a result, measuring the group velocity of the photoassociating field provides a direct measure of the pairing gap.

\section{Pairing and Group Velocity}

In order to account for the impact of Cooper pairing on the group velocity $v_g$ we include attractive pairing interactions into Eq.~(1) in the usual fashion via the Hamiltonian~ \cite{Meiser,Holland}
\begin{equation}
\label{HBCS} \hat{H}_{\rm BCS}=
\hat{H}-U\sum_{k,k'}\hat{s}_k^+\hat{s}_{k'}^-.
\end{equation}
The BCS ground state is found as usual by minimizing $\langle {\hat H}_{\rm BCS} - \mu {\hat N} \rangle$ , where $\mu$ is the chemical potential, using the ansatz
\begin{equation}
|{\rm BCS}\rangle = \prod_k(u_{\vec k} + v_{\vec k}{ \hat s}_{\vec k}^+)|0\rangle,
\end{equation}
with the result
\begin{equation}
\left(
\begin{array}{c}
u_{\vec k}^2\\
 v_{\vec k}^2
\end{array}
\right)
 =\frac{1}{2}\left (1\mp\frac{\xi_{\vec k}}{\sqrt{\xi_{\vec k}^2+|\Delta|^2}}\right )
\end{equation}
where $\eta_{\vec k}=\sqrt{\xi_{\vec k}^2+|\Delta|^2}$ is the mean-field quasiparticle energy, $\xi_{\vec k}=\epsilon_{\vec k}-\mu$ is the kinetic energy of the atoms measured from the Fermi surface, and
\begin{equation}
\Delta = U\sum_{\vec k} u_{\vec k}v_{\vec k}=\frac{U}{2} \sum_{\vec k}\frac{\Delta}{\sqrt{\xi_{\vec k}^2 + |\Delta|^2}}
\end{equation}
is the gap parameter.

The interaction Hamiltonian (\ref{HBCS}) does not modify the equations of motion for the operators $\hat m$, $\hat a$ and $\hat {\cal E}$. In the present context, its main effect in the weak excitation limit is to replace  $\langle {\hat S}_+{\hat S}_-\rangle$ by
\begin{equation}
\langle {\hat S}_+{\hat S}_-\rangle =\sum_\vec{k} v_\vec{k}^2
+\sum_{\vec{k}\neq \vec{k}'}
u_\vec{k}v_\vec{k}u_{\vec{k}'}v_{\vec{k}'} \simeq N +\left
(\frac{\Delta}{U} \right )^2.
\end{equation}

Within the weak-coupling limit of BCS theory, $\epsilon_\vec{k}$ and $\xi_{\vec k}$ are approximately independent of the wave vector ${\vec k}$, $\epsilon_\vec{k} \rightarrow\epsilon_F$ and $\xi_{\vec k} \rightarrow \xi$, where $\epsilon_F$ is the Fermi energy~ \cite{Ohashi}. In that case the group velocity becomes
\begin{equation}
v_{g, \Delta}=\frac{c}{1+\beta_\Delta},
\label{BCS group}
\end{equation}
where
\begin{equation}
\beta_f \rightarrow \beta_\Delta = \beta_f \left(1
+\frac{N\Delta^2}{4\xi^2+4\Delta^2} \right ),
\end{equation}
indicating that it now depends on both $N$ {\em and} the pairing gap
$\Delta$.
\begin{figure}[ht]
\includegraphics[width=0.9\columnwidth]{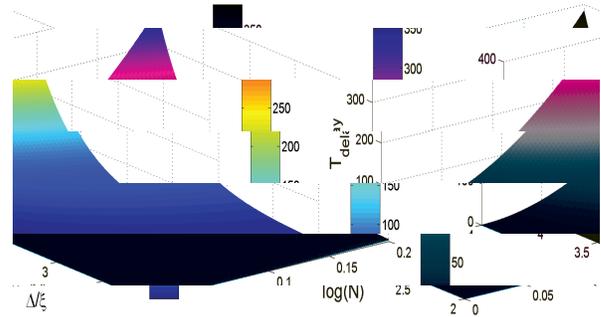}
\caption{Dimensionless relative time delay $T_d$ (scaled by $L/v_g$)
as a function of $N$ and the dimensionless pairing gap
$\Delta/\xi$.} \label{fig3}
\end{figure}
This is illustrated in Fig.~2, which shows the time delay
\begin{equation}
T_d=\frac{L}{v_{g, \Delta}}-\frac{L}{v_g}=\frac {L \beta_\Delta}{c}
\label{delay}
\end{equation}
experienced by a short photoassociating light pulse as a function of
$N$ and the pairing gap $\Delta$, relative to the delay in the
absence of gap. For large values of $\Delta$, we have
$v_{g,\Delta}\sim N^{-2}$, approaching the case of a bosonic
atom-molecule dark-state medium \cite{HJ}, with a gap-dependent
enhancement factor that is determined precisely by the ratio of the
molecule population $N_a(\Delta)$ and $N_a$ in the presence or
absence of a pairing gap,
\begin{equation}
\zeta=1+\frac{N\Delta^2}{4(\xi^2+\Delta^2)}=\frac{N_{a,\Delta}}{N_a},
\end{equation}
see Fig. ~3. That is, the variation in group velocity originates directly from the PA-induced atom-molecule superpositions in the $\Lambda$ level scheme of Fig.~1.
\begin{figure}[ht]
\includegraphics[width=0.95 \columnwidth]{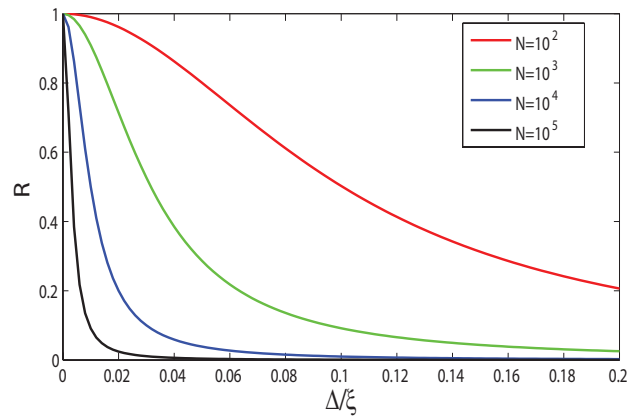}
\caption{Relative molecule population $\zeta^{-1}=N_a/N_a(\Delta)$
as a function of $N$ and the dimensionless paring gap $\Delta/\xi$.}
\label{fig3}
\end{figure}

\section{Conclusion}

In conclusion, we have shown that the two-color photoassociation of fermionic atoms into bosonic molecules via a dark-state transition results in a group velocity of the photoassociating field that can be slowed significantly, in complete analogy with the situation of EIT in lambda three-level atomic systems. That velocity $v_g$ depends not only on whether the atoms are bosonic or fermionic, with an associated $N^2$ versus $N$ dependence, but also on the possible pairing of the fermionic atoms resulting from attractive two-body interactions. As such, a measure of the propagation delay of the photoassociating light pulse ${\hat {\cal E}}(z,t)$ provides a direct measurement of the pairing gap $\Delta$. This nondestructive ${\it in~situ}$ diagnostic technique, which provides clear evidence of Fermi pairing in the weakly interacting BCS regime, supports and extends the idea of using Raman spectroscopy \cite{cote} to extract the pairing parameters, but differs from proposals based solely on the use of atomic transitions \cite{f-eit}.

In order to estimate the pairing-induced optical time delay of the propagating pulse, we consider the typical values $g\sim 100\mathrm{KHz}$, $\Omega\sim 1\mathrm{MHz}$, $N=10^5$, $L=1\mathrm{mm}$, and $\gamma_m\sim 16\mathrm{MHz}$, $\gamma_a\sim 600\mathrm{Hz}$ \cite{cote}. These values give for the bosonic sample a group velocity of $v_g\sim 3\mathrm{km}\cdot s^{-1}$, that is, a significant slowing down of the light pulse. For the normal Fermi gas, the significantly less favorable scaling of $v_g$ with $N$ instead of $N^2$ gives $v_g\sim 0.5c$, the rather small change that is expected to be challenging to observe. Finally, for paired fermionic atoms we find $v_g\sim 300\mathrm{km} \cdot {\rm s}^{-1}$ for $\Delta/\xi =0.2$, and $v_g\sim 15\mathrm{km} \cdot s^{-1}$ for $\Delta/\xi =2$, a change of two to three  orders of magnitude compared to the case of a normal Fermi gas. As already mentioned, for an increasing gap $v_g$ rapidly approaches the bosonic case. Note that shorter samples lead to a reduction in delay time $T_d$ that scales as $L^2$, as readily seen from Eqs.~(\ref{betaf}) and (\ref{delay}).

Our discussion ignores the decay of molecular states. However, it can be readily shown that after including these decay terms, the group velocity of the signal is still in the form of Eq.~(15), but with the substitution
$\Omega \rightarrow \sqrt{\Omega^2 +\gamma_m \gamma_a}$ ~\cite{HJ}.  In practice, the PA pulse duration $\tau$ should satisfy $ \tau \ll \gamma_a^{-1}\sim 1.67\mathrm{ms}$, a condition that can be
fulfilled in current experiments \cite{cote,d1,d2,d3}.

Future work will improve the sample description by incorporating its spatial profile in a more realistic multi-mode model, with a more detailed description of the two-body physics. In this context it will also be interesting to consider cavity-induced transparency with a degenerate Fermi gas \cite{Search}. A significantly more challenging problem will involve the situation of strong pair fluctuations at the BEC-BCS crossover \cite{Hu}. Finally, we note that the use of non-classical associating light fields may also allow one to consider the correlations of the transmitted field and/or a possible molecule-photon entanglement as probes of the Fermi pairing or perhaps of other exotic phases.

\begin{acknowledgments}
This work is supported by the U.S. National Science Foundation, by
the U.S. Army Research Office, and by the NSFC.
\end{acknowledgments}

\end{document}